\documentclass{article}
\usepackage{amssymb}
\usepackage{amsmath}

\newcommand{\ho}{\mathfrak{H}_0}
\newcommand{\hi}{\mathfrak{H}_\updownarrow}
\newcommand{\hl}{\mathfrak{H}(\lambda)}
\newcommand{\hlt}{\overline{\mathfrak{H}}(\lambda)}
\newcommand{\hm}{\mathfrak{H}_m}
\newcommand{\hn}{\mathfrak{H}_n}
\newcommand{\hh}{\check{H}}
\newcommand{\hho}{\check{H}_0}
\newcommand{\hhi}{\check{H}_\updownarrow}
\newcommand{\ag}{a_g^{}}
\newcommand{\agd}{a_g^\dagger}
\newcommand{\eng}{\hat{n}_g^{}}
\newcommand{\gi}{\lambda}
\newcommand{\hg}{\check{H}_g}
\newcommand{\app}{\ \overset{\ \lambda^2}{\simeq}\ }
\newcommand{\evol}{\mathfrak{E}}
\newcommand{\jc}{J\!\!C}
\newcommand{\jch}{\breve{J\!\!C}}
\newcommand{\ci}{\mathfrak{C}}

\begin{document}


\title{Trapped ions interacting with laser fields:
\\ a perturbative analysis without rotating wave approximation}
\author{P. Aniello$^{\ast}$, V. Man'ko$^\ddagger$, G. Marmo$^\ast$,
A. Porzio$^{\dagger}$, S. Solimeno$^{\dagger}$
\\
and F. Zaccaria$^\ast$
\\ \\
{\small Dipartimento di Scienze Fisiche, Universit\`a di Napoli ``Federico II''
} \\
{\small and} \\
{\small
$\ast$ Istituto Nazionale di Fisica Nucleare, Sezione di Napoli}
\\ {\small $\dagger $ ``Coherentia'' -- INFM, Unit\`{a} di Napoli}
\\
{\small $\ddagger$ Lebedev Physics Institute, Moscow}}

\maketitle


\begin{abstract}
The Hamiltonian describing a single ion placed in a potential trap
in interaction with a laser beam is studied by means of
a suitable perturbative approach. It is shown, in particular,
that the rotating wave approximation does not provide the correct expression,
already at the first perturbative order,
of the evolution operator of the system.
\end{abstract}


\section{Introduction} \label{introduction}

Trapped ions in interaction with laser beams are extremely useful
tools for investigating fundamental aspects of quantum physics.
For instance, they have been used for the generation of coherent,
squeezed~\cite{Meekhof} and Schr\"odinger-cat
states~\cite{Monroe}, and for the preparation of entangled Bell
and GHZ states~\cite{Turchette} \cite{Sorensen}. They have also
had important experimental applications as, for instance,
precision spectroscopy~\cite{Wineland} and laser
cooling~\cite{Wineland2} \cite{Bollinger} \cite{Vogt}.

Recently, the interest for laser-driven ion traps has received
a novel impulse in view of its applications in the fastly developing area of
quantum computing.
Indeed, in a quantum computer (QC),
information is stored in a `quantum register' composed of $N$
two-level systems representing the quantum bits, or qubits~\cite{Braunstein}.
The storage of data and all the basic operations are implemented
by inducing controlled dynamics on the quantum  register~\cite{Kilin}.
Since a QC is a quantum mechanical system, it can perform
superpositions of computation operations with a remarkable
gain of efficiency with respect to a classical computer; a typical example
is the solution of the problem, fundamental for some cryptographic schemes,
of factoring large numbers into primes~\cite{Shor}.
In 1995 Cirac and Zoller~\cite{Cirac}
proposed a concrete model for a ion-trap computer consisting
of $N$ atomic ions trapped in a parabolic potential well,
each ion being regarded as a two-level system, hence as a realization
of a qubit.
The control of the quantum degrees of freedom is achieved by
addressing the ions with time, frequency and intensity
controlled laser beams (see also~\cite{Pellizzari}~\cite{Plenio}).

All the applications mentioned above are realized by ion traps with a
linear geometry. In such traps a strong confinement is induced along the
$y$ and $z$ axes, while a weak harmonic confinement is induced along the
principal trap axis $x$.
This confinement scheme is realized by the Paul linear
trap~\cite{Wineland3}. $N$ ions in a linear trap will form
a chain along the principal trap axis. This allows to reduce the initial
$3N$-dimensional model to a $N$-dimensional one,
which can be treated conveniently by the introduction
of the normal coordinates of the ion chain~\cite{James}.
As already mentioned,
controlled dynamics can be induced on the ion trap by means of
laser beams. The Hamiltonian describing the total
system~\cite{James}~\cite{Cirac2} --- which we will
call simply the `ion trap Hamiltonian' (ITH) --- in spite of its relative
formal simplicity, gives rise to a Schr\"odinger equation whose exact
solutions are not known and its study requires the
adoption of suitable approximations. The stantardly used method is
the rotating wave approximation
(RWA)~\cite{Allen} \cite{Schleich}.
The RWA is a very popular technique in quantum optics and, in general, in
the study of resonance phenomena since it leads to considerable
simplifications in many calculation procedures. It consists
essentially in passing to the interaction picture
and then dropping those terms of the effective Hamiltonian which are
rapidly oscillating (usually called `counter rotating terms'
or `virtual terms'). Using this technique
and the Lamb-Dicke approximation~\cite{Jonathan}, the ITH can be reduced
to an effective Hamiltonian
formally identical to the Hamiltonian of the Jaynes-Cummings model
(JCM)~\cite{Jaynes} \cite{Shore} in the interaction picture,
hence explicitly integrable.

Despite its great popularity, the {\it general validity}
of the RWA is rather uncertain. In particular,
the application of the RWA to the evaluation of the evolution operator
is justified only by semiqualitative arguments
and the ubiquity of this procedure is mainly due to the
chance of performing explicit calculations partially supported by the
prediction of some experimentally observable phenomena, for instance
the typical `collapses and revivals'
in two-level systems~\cite{Eberly} \cite{Rempe} \cite{Fleischhauer}.
Thus, some attempts of taking into account the impact of counter rotating
terms have been made. Specifically, perturbative corrections
to the energy spectrum~\cite{Cohen} and corrections to the time evolution
by means of path integral~\cite{Zaheer} and
perturbative~\cite{Vyas} \cite{Phoenix}
\cite{Fang} techniques have been investigated.
It should be also mentioned that an attempt of considering the counter
rotating contributions is already present in the classical study
of the magnetic resonance done by Bloch and Siegert~\cite{Bloch} and
in the later related work of Shirley~\cite{Shirley}. Anyway,
the validity of all these approaches,
as well as of the RWA itself~\cite{Tavis}, rests on the smallness
of the coupling constant, which in the case of a laser-driven
ion trap is proportional to the intensity of the laser field.
This is a severe drawback
since an intense laser field
implies a fast coupling between the two internal
energy levels of the trapped ions i.e., for instance, a fast QC.

In the present paper, we propose to study the ITH using a perturbative
approach along new lines with respect to the exhisting literature.
Since, by virtue of the normal coordinates of the ion chain, the
$N$-ion case does not introduce any essential complication with
respect to the single ion case (especially if the nonharmonic
component of the ion-ion interaction can be neglected), we will
restrict, for the sake of simplicity,
to the latter case. Obviously, the $N$-ion case is of
great interest for the applications and will deserve a particular
attention in a forthcoming paper~\cite{Aniello1}.
We proceed as follows. First, we notice that passing to a
`rotating frame', i.e.\ to a suitable interaction picture, the
ITH transforms into a time-independent Hamiltonian which we will call
the `rotating frame Hamiltonian' (RFH). At this point, we use the fact that
--- as it has been shown by some of the authors~\cite{Aniello0} ---
the RFH is unitarily equivalent to a Hamiltonian formally similar to the RFH,
except for the fact that the new coupling constant is not proportional
any more to the field intensity but a simple bounded function of it
(see section~{\ref{ion}}).
We will call this Hamiltonian the `balanced Hamiltonian' (BH).
The BH is an ideal starting point for a perturbative approach,
since the results obtained by its study
hold also in the strong field regime.
Next, using the tools of perturbation
theory for linear operators~\cite{Kato} \cite{Reed},
we develop a perturbative procedure which allows to give approximate
expressions of the evolution operator associated with the BH (hence with
the ITH) in terms of unitary operators.
Our basic idea is the following. The Jaynes-Cummings Hamiltonian is
exactly solvable --- i.e.\ its spectrum is discrete and its
eigenvalues and eigenvectors are known ---
by virtue of the fact that it can be written as the
sum of two commuting operators and one of the two is trivially
solvable (see section~{\ref{classical}}).
Then, in general, we wonder if, given a
perturbed Hamiltonian $\mathfrak{H}(\lambda)=\ho+\lambda\,\hi$,
with $\ho$ exactly solvable,
we can build a procedure which allows to compute, for each
$n=1,2,\ldots$, hermitian operators
$\ho^{(n)}(\lambda)$, $\ci^{(n)}(\lambda)$
depending analytically on the parameter $\lambda$, such that
\begin{equation} \label{uno}
\mathfrak{H}(\lambda) = \ho^{(n)}(\lambda) +
\ci^{(n)}(\lambda) +\mathcal{O}(\lambda^{n+1}),
\end{equation}
$\ho^{(n)}(\lambda)$ is exactly solvable and
\begin{equation} \label{due}
\left[\ci^{(n)}(\lambda),\, \ho^{(n)}(\lambda)\right]=0 \, .
\end{equation}
Indeed, in this case the determination of an approximate expression,
at each perturbative order, of the evolution operator associated with
$\mathfrak{H}(\lambda)$ is greatly simplified.
We will show in section~{\ref{analysis}} that this idea is correct and
we will apply the method to the BH in section~{\ref{treatment}}.
In sections~{\ref{spectrum}} and~{\ref{t-evolution}},
we compare the perturbative expressions obtained for the energy spectrum
and the evolution operator with the results obtained applying
the RWA to the BH.
It will be shown that, while the RWA gives the right first order
correction to the unperturbed eigenvalues,
it does not give the right first order expression for the evolution operator
since it neglects corrections to the eigenprojectors.
Eventually, in section~{\ref{discussion}}, conclusions are drawn.


\section{The Jaynes-Cummings model} \label{classical}

In this section we will give a very concise treatment of the classical
JCM. This will allow us to gain a better insight in our perturbative
analysis of the ion trap Hamiltonian.

The Hamiltonian of the JCM reads ($\hbar=1$):
\begin{equation}
H_\mathrm{JC} = \nu\,\hat{n}+\frac{1}{2}\omega\,\sigma_z+\gi\,\nu\left(
a\,\sigma_+ +a^\dagger\,\sigma_-\right),
\end{equation}
where $a$, $a^\dagger$, $\hat{n}$ are the annihilation, creation and number
operators and $\sigma_z$, $\sigma_\pm$ the Pauli operators.
In the following, we will denote by $\{|n\rangle\}$ the Fock basis and by
$\left|g\right\rangle, \left|e\right\rangle$ the eigenvectors of
$\sigma_z$:
\[
\hat{n} \left|n\right\rangle = n \left|n\right\rangle,\ \ \
\sigma_z \left|g\right\rangle=-\left|g\right\rangle, \
\sigma_z\left|e\right\rangle=\left|e\right\rangle.
\]
The JCM is exactly solvable. This is due to the fact that its Hamiltonian
can be easily represented as the sum of two constants of the motion.
Indeeed, observe that $H_\mathrm{JC}$ can be written as
$H_\mathrm{JC}=\mathcal{N}+\mathcal{S}$, where
\begin{equation}
\mathcal{N} :=\nu\left(\hat{n}+\frac{1}{2}\sigma_z\right),\ \ \
\mathcal{S} :=\frac{1}{2}\left(\omega-\nu\right)\sigma_z+\gi\,\nu
\left(a\,\sigma_+ +a^\dagger\,\sigma_-\right)
\end{equation}
and
\begin{equation}
[\mathcal{N},\mathcal{S}]=[\mathcal{N},H_\mathrm{JC}]=
[\mathcal{S},H_\mathrm{JC}]=0.
\end{equation}
It follows that the evolution operator of the JCM factorizes as
\[
e^{-i H_\mathrm{JC} t}=
e^{-i \mathcal{N} t}\, e^{-i \mathcal{S} t}.
\]
Moreover, $\mathcal{N}$ has a discrete spectrum and its eigenspaces, namely
the one-dimensional eigenspace
$
\mathcal{H}_0=\mathrm{Span}\{|0\rangle\otimes|g\rangle\}
$
and the two-dimensional eigenspaces
\[
\mathcal{H}_{n}=\mathrm{Span}\{|n-1\rangle\otimes|e\rangle, \ \
|n\rangle\otimes|g\rangle\},\ \ \ n=1,2,\ldots\ ,
\]
are invariant subspaces for $\mathcal{S}$,
which can be diagonalized in each of
these mutually orthogonal subspaces. This yelds to a diagonalization of the
whole Hamiltonian and to an explicit expression for the unitary operator
$e^{-i \mathcal{S} t}$.
If the {\it resonance condition} $\nu=\omega$ is satisfied, then this
operator assumes a particularly simple form and can be also computed
by direct exponentiation. indeed, using the fact that
\begin{eqnarray*}
\left(a\,\sigma_+ + a^\dagger\,\sigma_-\right)^{2m}
\!\! & = & \!\!
\left(a\, a^\dagger\, \sigma_+\,\sigma_- + a^\dagger a\, \sigma_-\,\sigma_+
\right)^m
\\
& = & \!\!
\hat{n}^m \left|g\right\rangle\left\langle g\right| +
\left(\hat{n}+1\right)^m \left|e\right\rangle\left\langle e\right| ,
\end{eqnarray*}
we find:
\begin{eqnarray}
\jc(t)
\!\! & := & \!\!
\exp\left(-i\mathcal{S}\, t\right)
\nonumber \\
& = & \!\!
\cos\left(\gi\,\nu\,\sqrt{\hat{n}}\ t\right)
\left|g\right\rangle\left\langle g\right| +
\cos\left(\gi\,\nu\,\sqrt{\hat{n}+1}\ t\right)
\left|e\right\rangle\left\langle e\right|
\nonumber \\
& - & \!\!
i\left(\dfrac{\sin\left(\gi\,\nu\,\sqrt{\hat{n}+1}\ t\right)
}{\sqrt{\hat{n+1}}}\, a\, \sigma_+
+\,\dfrac{\sin\left(\gi\,\nu\,\sqrt{\hat{n}}\ t\right)
}{\sqrt{\hat{n}}}\, a^\dagger\, \sigma_-\right) .
\end{eqnarray}
We will call $\jc(t)$ the {\it Jaynes-Cummings evolutor}.

We conclude this section observing that since
\[
e^{-i\frac{\pi}{2}\hat{n}}\, a\, e^{i\frac{\pi}{2}\hat{n}}=i\, a,
\]
where $e^{i\frac{\pi}{2}\hat{n}}$ is nothing but the Fourier-Plancherel
operator, the Hamiltonian $H_\mathrm{JC}$ is unitarily equivalent to
the following one:
\begin{equation}
\widehat{H}_\mathrm{JC} := e^{-i\frac{\pi}{2}\hat{n}}\,
H_\mathrm{JC}\, e^{i\frac{\pi}{2}\,\hat{n}}=
\mathcal{N} + \widehat{\mathcal{S}},
\end{equation}
where
\begin{equation}
\widehat{\mathcal{S}}:= \frac{1}{2}\left(\omega-\nu\right)\sigma_z +
i\,\gi\,\nu\left(a\,\sigma_+ - a^\dagger\,\sigma_-\right).
\end{equation}


\section{The ion trap Hamiltonian} \label{ion}

A two-level ion of mass $\mu$ in a potential trap,
with strong confinement along the $y$ and $z$ axes, and weak harmonic
binding of frequency $\nu$ along the $x$-axis (the `trap axis'),
can be described --- neglecting the motion of the ions transverse to the
trap axis --- by a Hamiltonian of the following type
($\hbar=1$):
\[
H_0=\nu \, a^{\dagger} a^{}+\frac{1}{2}\omega_{ge}\,\sigma_z,
\]
where $a$ is the vibrational annihilation operator
\[
a=\left(\frac{\mu\,\nu}{2}\right)^{\frac{1}{2}}
\left(\hat{x}+\frac{i}{\mu\,\nu}\,\hat{p}_x\right)
\]
and $\sigma_z$ the effective spin operator
associated with the internal degrees of freedom of the ion.
Let us suppose now that
the ion is addressed by a laser
beam of frequency $\omega _{L}$ in a traveling wave configuration.
Then, the Hamiltonian describing the physical system (ITH) becomes
\begin{equation}
H(t)=H_0 + H_\updownarrow(t),
\end{equation}
where
\begin{equation}
H_\updownarrow (t) :=
\Omega_{R}\left( e^{i\omega_{L}t}\,\sigma_{-}\,
D(i\eta)^{\dagger}+
e^{-i\omega_{L}t}\,\sigma _{+}\, D(i\eta)\right),
\end{equation}
with
$\Omega_{R}=\wp\,\mathcal{E}$ the Rabi frequency and $\mathcal{E}$
the intensity of the laser field.
Moreover, we have set:
\begin{equation}
D(i\eta):=\exp \left(i \eta \left(
a + a^{\dagger}\right) \right),
\end{equation}
where
\begin{equation}
\eta := \frac{k_L\cos\phi}{\sqrt{2\mu\nu}}
\end{equation}
--- with $\mathbf{k}_L$ the wavevector and $\phi$ the angle between the
$x$-axis and $\mathbf{k_L}$ ---
is the Lamb-Dicke factor. Notice that
$D(\alpha)$, $\alpha\in\mathbb{C}$,
is a displacement operator, namely
\begin{equation}
D(\alpha):=\exp \left( \alpha\, a^{\dagger}-
\alpha^*  a \right),\ \ \
D(\alpha)\, a\, D(\alpha)^\dagger = a -\alpha.
\end{equation}
In order to work with operator matrices,
the Hilbert space of the total system (`pointlike' ion + internal degrees
of freedom of the ion), namely $\mathcal{H}\otimes\mathbb{C}^2,\
\mathcal{H}\equiv L^2(\mathbb{R})$, will be
identified with the space $\mathcal{H}\oplus\mathcal{H}$.

Now, the dynamical problem associated with the time-dependent Hamiltonian
$H(t)$ can be turned into a time-independent
problem. Indeed, switching to the interaction picture with
reference Hamiltonian $\frac{1}{2}\omega_Lt\,\sigma_{z}$ and setting
\begin{equation}
R_t^{} := \exp\left(i\frac{1}{2}\omega_Lt\,\sigma_z \right),
\end{equation}
one obtains the time-independent `rotating frame Hamiltonian' (RFH)
\begin{eqnarray}
\widetilde{H} \!\! &=&\!\! R_{t}^{}\left(H(t)-\frac{1}{2}
\omega _{L}\,\sigma_{z}\right)
R_{t}^{\dagger }  \nonumber \\
&=& \!\!
\nu\,\hat{n} +\frac{1}{2}\delta\,
\sigma_{z}+\Omega_{R}\left( \sigma_{-}\, D(i\eta)^{\dagger}+\sigma_{+}\,
D(i\eta) \right) ,
\end{eqnarray}
where $\hat{n}=a^\dagger a$ is the number operator and
$\delta:=\omega_{ge}-\omega_L$ is the ion-laser detuning.\\
At this point, in order to give an approximate expression
of the evolution operator associated with the Hamiltonian $\widetilde{H}$,
the {\it rotating wave approximation} (RWA) is usually applied.
It consists in
expanding the exponential $D(i\eta)$,
then passing to the interaction picture with reference Hamiltonian
$\widetilde{H}_0=\nu\,\hat{n} +\frac{1}{2}\delta\,\sigma_z,$
so obtaining the interaction picture Hamiltonian
\[
\widetilde{H}_{\mathrm{int}}(t)= \Omega_R
\left(1-i\eta
(e^{i\nu t}\, a^\dagger + e^{-i\nu t}\, a)+\ldots\right)
e^{-i\delta t}\,\sigma_- + h.c.\ ,
\]
and, finally, retaining only that terms in $\widetilde{H}_{\mathrm{int}}(t)$
which are slowly rotating.
Hence --- assuming that $\eta\ll 1$ (Lamb-Dicke regime), so that
one can keep only the terms which are at most linear in $\eta$ ---
in correspondence to the three types of resonance
\begin{equation} \label{res}
\delta=\omega_{ge}-\omega_L\simeq 0,\ \ \
\delta +\nu\simeq 0,\ \ \
\delta -\nu\simeq 0,
\end{equation}
one obtains respectively the following three types of effective
interaction picture Hamiltonian:
\[
\widetilde{H}_{\mathrm{eff}}^{(0)}
=\Omega_R\left(\sigma_-+\sigma_+\right),
\]
\[
\widetilde{H}_{\mathrm{eff}}^{(-)}=
i\eta\,\Omega_R\left(a^\dagger\,\sigma_+-
a\,\sigma_-\right),\ \ \
\widetilde{H}_\mathrm{eff}^{(+)}=
i\eta\,\Omega_R\left(a\,\sigma_+
- a^\dagger\,\sigma_-\right).
\]
These effective Hamiltonians, in correspondence to the respective resonances,
commute with the reference Hamiltonian $\widetilde{H}_0$.
This is due to the fact that the resonances~{(\ref{res})} are
associated with degeneracies of the reference Hamiltonian.
In fact, it turns out that
the spectrum of $\widetilde{H}_0$ is degenerate if and only if the
condition $m\,\nu=|\delta|,\ m=0,1,2,\ldots$, holds. Notice that, in particular,
$\widetilde{H}_\mathrm{eff}^{(+)}$ is equal ---
up to a unitary transformation and setting $\delta=\omega$ and
$\eta\,\Omega_R=\gi\,\nu$ --- to the constant of the motion $\mathcal{S}$
of the Jaynes-Cummings Hamiltonian in the resonant regime
(see section~{\ref{classical}}).

It will be shown that the RWA is a rather poor approximation
for the evaluation of the evolution operator.
The argument usually adopted in order to support its
validity is the following. Let us consider the Feynman-Dyson expansion
of the evolution operator $U_\mathrm{int}(t,t_0)$
associated with the interaction picture Hamiltonian
$\widetilde{H}_{\mathrm{int}}(t)$. At the first order one has:
\[
U_\mathrm{int}(t,t_0)\simeq
\mathrm{Id}-i\int_{t_0}^t \widetilde{H}_\mathrm{int}(\tau)\ d\tau .
\]
Then, it is argued that
the fastly oscillating terms in $\widetilde{H}_\mathrm{int}$
give a smaller contribution to the integral in the r.h.s.\ of the
previous formula with respect to the slowly rotating ones.
As it will be seen later on, this argument turns out to be erroneous.
The misunderstanding stems from the fact that
one is using a perturbative expansion
of the evolution operator whose terms (except the identity)
are not unitary operators.\\
Besides, any perturbative approach
does not work in the strong field regime, since the
Rabi frequency $\Omega_R$ which appears in the
interaction component of the RFH is proportional to the
intensity of the laser field. This problem can be bypassed
by means of a suitable unitary transformation
of the Hamiltonian $\widetilde{H}$ (see~\cite{Aniello0}).
This transformation allows to obtain a
`balanced Hamiltonian' (BH) $\breve{H}$
which is the sum of a large component
having a simple `diagonal' form --- i.e. such that its matrix representation
in the standard basis
\begin{equation} \label{standard}
\!\!\!\{
|n\rangle\otimes|g\rangle,\
|n\rangle\otimes|e\rangle :\
n = 0,1,2,\ldots\}
\end{equation}
is diagonal --- and a small interaction component scarcely sensitive to the
Rabi frequency $\Omega_R$.
Indeed, introduced the dimensionless parameter $\Delta :=\delta/\Omega_R$
(from this point onwards we will assume that $\Omega_R\neq 0$),
there is a unitary operator $T_\Delta$ such that
\begin{equation}
\breve{H}:= T_\Delta^{}\, \widetilde{H}\, T_\Delta^\dagger
= \breve{H}_0 +\breve{H}_\updownarrow ,
\end{equation}
where we have set:
\begin{equation}
\breve{H}_0 := \nu\,\hat{n} +
\frac{1}{2}\breve{\delta}\,\sigma_z + \gi\,\breve{\eta}\, \nu ,
\end{equation}
with
\begin{equation}
\breve{\delta}:= \sqrt{4\Omega_R^2+\delta^2},\ \ \
\breve{\eta}:=\frac{\Delta}{\sqrt{4+\Delta^2}}\,\eta,\ \ \
\gi := \frac{1}{\sqrt{4+\Delta^2}}\,\eta
=\frac{1}{\Delta}\,\breve{\eta},
\end{equation}
and
\begin{eqnarray}
\breve{H}_\updownarrow
\!\! & := & \!\!
i\,\gi\,\nu\left(
\left(a - a^\dagger\right)
\left(\sigma_+\, D(i\breve{\eta})
+\sigma_-\, D(i\breve{\eta})^\dagger\right)
- i\breve{\eta}
\left(\sigma_+\, D(i\breve{\eta})
-\sigma_-\, D(i\breve{\eta})^\dagger\right)\right)
\nonumber \\
\!\! & = & \!\!
i\,\gi\,\nu\left(a-a^\dagger\right)\left(\sigma_+
+\sigma_-\right)
\\
\!\! & + & \!\!
i\,\gi\,\nu
\sum_{m=1}^\infty \frac{(i\breve{\eta})^{m}}{m!}\left(a^2-a^{\dagger\,2}+1
-m\right)(a+a^\dagger)^{m-1}\left(\sigma_+ +(-1)^{m}\sigma_-\right).
\nonumber
\end{eqnarray}
Notice that the `balanced detuning' $\breve{\delta}$ is, unlike $\delta$,
always positive so that the degeneracy condition for $\breve{H}_0$ is
now: $\nu=m\,\breve{\delta},\ m=1,2,\ldots\ $.\\
The unitary operator $T_\Delta$ has the following explicit form:
\begin{eqnarray}
\!\!\! T_\Delta^{}
\!\! & = & \!\!\!\!\!
\left[\!\!
\begin{array}{cl}
\varkappa_\Delta^+\,
D(i\epsilon_\Delta^-\,\eta)
& \varkappa_\Delta^-\,
D(i\epsilon_\Delta^+\,\eta)
\\
&
\\
-\varkappa_\Delta^-\,
D(i\epsilon_\Delta^+\,\eta)^\dagger
& \varkappa_\Delta^+\,
D(i\epsilon_\Delta^-\,\eta)^\dagger
\end{array}
\!\!\right]
\nonumber \\
& &
\\
\!\! & = & \!\!\!\!\!
\left[\!\!
\begin{array}{cl}
\varkappa_\Delta^+\,
D\left(i(\breve{\eta}-\eta)/2\right)
& \varkappa_\Delta^-\,
D\left(i(\breve{\eta}+\eta)/2\right)
\\
&
\\
-\varkappa_\Delta^-\,
D\left(i(\breve{\eta}+\eta)/2\right)^\dagger
& \varkappa_\Delta^+\,
D\left(i(\breve{\eta}-\eta)/2\right)^\dagger
\end{array}
\!\!\right], \nonumber
\end{eqnarray}
with
\begin{equation}
\varkappa_\Delta^\pm =
\sqrt{\frac{1}{4}+\frac{1}{2\sqrt{4+\Delta^2}}}\, \pm \,
\mathrm{sign}(\Delta)\,  \sqrt{\frac{1}{4}-\frac{1}{2\sqrt{4+\Delta^2}}}\ ,
\end{equation}
\begin{equation}
\epsilon_\Delta^\pm =
\frac{\Delta}{2\sqrt{4+\Delta^2}}\,\pm\,\frac{1}{2}\ .
\end{equation}
The unitary operator $T_\Delta$ can be decomposed as the product
of three unitary transformations:
\[
T_\Delta=T_3\,T_2\,T_1.
\]
The transformation $T_1$ has been introduced
by Moya-Cessa {\it et al.}~\cite{Moya} and has the following form:
\begin{eqnarray}
T_1^{} \!\! & := & \!\!
\frac{1}{\sqrt{2}}\left(\frac{1}{2}\left(\mathcal{D}+
\mathcal{D}^\dagger\right)-
\frac{1}{2}\left(\mathcal{D}-\mathcal{D}^\dagger\right)
\sigma_z+\mathcal{D}\,\sigma_+
-\mathcal{D}^\dagger\,\sigma_-\right)
\nonumber \\
\!\! & = & \!\!
\frac{1}{\sqrt{2}}\left[
\begin{array}{rl}
\mathcal{D}^{\dagger } & \mathcal{D} \\
-\mathcal{D}^{\dagger } & \mathcal{D}
\end{array}
\right] ,\ \ \ \ \
\mathcal{D}\equiv D(i\eta/2).
\nonumber
\end{eqnarray}
The transformation
$T_2^{}$ is a spin rotation by the angle $\theta$ round the
$y$-axis:
\[
T_{2}^{} :=\left[
\begin{array}{lr}
\cos \theta/2 & -\sin \theta/2 \\
\sin \theta/2 & \cos \theta/2
\end{array}
\right] ,
\]
where the angle $\theta$, $-\pi/2\le\theta\le\pi/2$, verifies the condition
\[
\tan\theta = \frac{\Delta}{2}=\frac{\omega_{ge}-\omega_L}{2\Omega_R}.
\]
The third transformation is given by
\begin{eqnarray}
T_3 \!\! & := & \!\! \frac{1}{2}\left(D(i\breve{\eta}/2)+
D(i\breve{\eta}/2)^\dagger\right)+
\frac{1}{2}\left(D(i\breve{\eta}/2)-
D(i\breve{\eta}/2)^\dagger\right)\sigma_z
\nonumber \\
\!\! & = & \!\!
\left[
\begin{array}{cc}
D(i\breve{\eta}/2 ) & 0
\\
0 & D(i\breve{\eta}/2)^\dagger
\end{array}
\right].
\nonumber
\end{eqnarray}
One can check that in the weak field limit $\Omega_R\rightarrow 0$
($\Delta=\delta/\Omega_R\rightarrow\pm\infty$) and in the strong field limit
$\Omega_R\rightarrow+\infty$ ($\Delta\rightarrow 0$)
the transformation $T_\Delta$ has the following behaviour:
\[
\lim_{\Delta\rightarrow +\infty} T_\Delta=\mathrm{Id},\ \ \
\lim_{\Delta\rightarrow -\infty} T_\Delta= T_2(-\pi)
=\left[\begin{array}{rr} 0 & 1 \\
-1 & 0
\end{array}\right],
\ \ \ \ \ \ \ (\mbox{weak field limit});
\]
\[
\lim_{\Delta\rightarrow 0} T_\Delta= T_1 \ \ \ \ \ \ \ \ \
(\mbox{strong field limit}).
\]
Thus, in the weak field limit, $T_\Delta$ goes to the identity if $\delta>0$
and to a spin rotation which sends $\sigma_z$ into $-\sigma_z$
for $\delta<0$ (so that, as already observed, $\breve{\delta}$ is
always positive); while, in the strong field limit, it goes to the transformation
introduced by Moya-Cessa {\it et alii}.

The constant $\gi$ which appears in the expression of the interaction
component $\breve{H}_\updownarrow$ of the balanced Hamiltonian
$\breve{H}$ will play the role of
perturbative parameter in our later analysis.
Moreover, it will
be seen that the constant $\breve{\eta}$ rules the number of terms that
have to be considered at each perturbative order.
We remark that both $\gi$ and $\breve{\eta}$ are bounded functions
of the Rabi frequency $\Omega_R$; indeed:
\[
0\le |\gi|=\frac{\Omega_R}{\sqrt{4\Omega_R^2+\delta^2}}\,|\eta|
\le\frac{1}{2}\,|\eta|
\]
and
\[
0\le|\breve{\eta}|=\frac{|\delta|}{\sqrt{4\Omega_R^2+\delta^2}}\,
|\eta|\le|\eta|.
\]
Thus, one can apply a perturbative approach also in the case of a large
Rabi frequency (hence in presence of an intense laser field).

\section{Perturbative analysis: outline of the method} \label{analysis}

Let $\ho$, $\hi$ be hermitian operators
and assume that $\ho$ has a purely discrete spectrum (i.e.\ it has a
pure point spectrum with finite-dimensional eigenspaces).
Denote by
\[
E_0 < E_1 < E_2 < \ldots
\]
the (possibly degenerate) eigenvalues of $\ho$ and by $P_0, P_1, P_2,\ldots$
the associated eigenprojectors.\\
Now, consider the operator
\[
\hl=\ho+\lambda\,\hi \ \ \ \ \lambda\in\mathbb{C},
\]
which is hermitian if $\lambda$ is real. It is possible to show that,
under certain conditions (see~{\cite{Kato}}~{\cite{Reed}}),
there exist positive constants $r_0,r_1,r_2,\ldots$
and a simply connected neighbourhood
$\mathcal{I}$ of zero in $\mathbb{C}$ such that the following
contour integral on the complex plane
\begin{equation}
P_m(\lambda)=\frac{i}{2\pi}\oint_{|E-E_m|=r_m}\!\!\!\!\!\!\!\!\!\!\!\!\!\!
\!\!\!\!\!\!\!\!\!\!
dE\ \ \ \ \ \left(\hl-E\right)^{-1}  \ \ \ \
\lambda\in\mathcal{I},
\end{equation}
defines a projection ($P_m(\lambda)^2=P_m(\lambda)$), which is an
orthogonal projection for real $\lambda$, with $P_m(0)=P_m$, and
$\mathcal{I}\ni\lambda\mapsto P_m(\lambda)$
is an analytic operator-valued function. Moreover, the range of
$P_m(\lambda)$ is an invariant subspace for $\hl$, hence
\begin{equation} \label{inv}
\hl\, P_m(\lambda) = P_m(\lambda)\, \hl\, P_m(\lambda),
\end{equation}
and there exists an analytic family $U(\lambda)$
of invertible operators such that
\begin{equation} \label{trasf}
P_m = U(\lambda)\, P_m(\lambda) \, U(\lambda)^{-1},\ \ \
U(0)=\mathrm{Id},
\end{equation}
and
\begin{equation}
U(\lambda)=e^{iZ(\lambda)}\ \ \ \lambda\in\mathcal{I},
\end{equation}
with $Z(\lambda^*)=Z(\lambda)^\dagger$ (hence, for real $\lambda$,
$Z(\lambda)$ is hermitian and $U(\lambda)$ is unitary), where
$\mathcal{I}\ni\lambda\mapsto Z(\lambda)$ is analytic.
Let us observe explicitly that the following relation holds:
\begin{equation} \label{rel}
U(\lambda^*)=e^{iZ(\lambda^*)}=e^{iZ(\lambda)^\dagger}=
U(\lambda)^{-1\,\dagger}.
\end{equation}
We remark that, as it is easily shown, the function
$\lambda\mapsto U(\lambda)$ is not defined uniquely by
condition~{(\ref{trasf})} even in the simplest case when
$\ho$ has a nondegenerate spectrum.\\
Now, let us define the operator $\hlt$ by
\begin{equation}
\hlt:=U(\lambda)\, \hl\, U(\lambda)^{-1},
\end{equation}
which, for real $\lambda$, is unitarily equivalent to $\hl$.
Using relations~{(\ref{inv})} and (\ref{trasf}), we find
\begin{eqnarray*}
\hlt\, P_m
\!\! & = & \!\!
U(\lambda)\, \hl\, P_m(\lambda)\, U(\lambda)^{-1}
\\
\!\! & = & \!\!
U(\lambda)\, P_m(\lambda)\,\hl\,P_m(\lambda)\, U(\lambda)^{-1}
\end{eqnarray*}
and hence:
\begin{equation}
\hlt\, P_m= P_m \,\hlt\, P_m \ \ \ \ m=0,1,\ldots\ .
\end{equation}
It follows that
\begin{equation}
\left[\ho,\,\hlt\right]=0
\end{equation}
and then we obtain the following important decomposition formula
\begin{equation} \label{decomposition}
U(\lambda)\, \hl\, U(\lambda)^{-1} = \ho + C(\lambda),
\end{equation}
where $\left[C(\lambda),\ho\right]=0$, i.e.\ $C(\lambda)$ is a constant of
the motion for the evolution generated by $\ho$. Notice that by virtue of
relation~{(\ref{rel})} we have:
\begin{eqnarray*}
C(\lambda^*)
\!\! & = & \!\!
U(\lambda^*)\, \mathfrak{H}(\lambda^*)\, U(\lambda^*)^{-1} - \ho \\
\!\! & = & \!\!
U(\lambda)^{-1\,\dagger}\,\hl^\dagger\, U(\lambda)^\dagger -\ho \\
\!\! & = & \!\!
C(\lambda)^\dagger.
\end{eqnarray*}
Thus, in particular, for real $\lambda$, $C(\lambda)$ is hermitian.

At this point, we are ready to build the perturbative decomposition
of the operator $\hl$ anticipated in section~{\ref{introduction}}.
Indeed, since the functions
$\lambda\mapsto C(\lambda)$ and $\lambda\mapsto Z(\lambda)$ are analytic
in $\mathcal{I}$ and $C(0)=Z(0)=0$, we can write:
\begin{equation} \label{power}
C(\lambda)=\sum_{m=1}^{\infty}\lambda^m\, C_m,\ \ \
Z(\lambda)=\sum_{m=1}^{\infty}\lambda^m\, Z_m \ \ \ \
\lambda\in\mathcal{I};
\end{equation}
here we notice that --- as $C(\lambda^*)=C(\lambda)^\dagger$ and
$Z(\lambda^*)=Z(\lambda)^\dagger$ --- for any $m$, $C_m=C_m^\dagger$
and $Z_m=Z_m^\dagger$. Then, if we set
\begin{eqnarray} \label{forhon}
\ho^{(n)}(\lambda)
\!\! & := & \!\!
e^{-i(\lambda\,Z_1+\cdots +\lambda^n Z_n)}\,
\ho \, e^{i(\lambda\,Z_1+\cdots+\lambda^n Z_n)},
\\ \label{forcin}
\ci^{(n)}(\lambda)
\!\! & := & \!\!
e^{-i(\lambda\, Z_1+\cdots +\lambda^n Z_n)}
\left(\lambda\, C_1+\cdots +\lambda^n C_n\right)\,
e^{i(\lambda\,Z_1+\cdots +\lambda^n Z_n)},
\end{eqnarray}
by virtue of formula~{(\ref{decomposition})}
we find that equation~{(\ref{uno})}
is satisfied together with condition~{(\ref{due})}.\\
Now, in order to determine the operators $\{C_n(\lambda)\}$ and
$\{Z_n(\lambda)\}$, let us substitute the exponential form $e^{iZ(\lambda)}$
of $U(\lambda)$ in formula~{(\ref{decomposition})}; we obtain:
\begin{equation}
\mathfrak{H}(\lambda) + \sum_{m=1}^{\infty} \frac{i^m}{m!}\,
\mathrm{ad}_{Z(\lambda)}^m\, \mathfrak{H}(\lambda)= \ho+ C(\lambda),
\end{equation}
where we recall that $\mathrm{ad}_K\, L = [K,L]$.
Next, substituting the power expansions~{(\ref{power})} in this equation,
in correspondence to the various orders in the
perurbative parameter $\lambda$, we get the following set of conditions:
\begin{eqnarray}
C_1 -i\left[Z_1,\ho\right]-\hi=0
& , &
\left[C_1,\ho\right]=0 \label{first}
\\
C_2-i\left[Z_2,\ho\right]+\frac{1}{2}\left[Z_1,\left[Z_1,\ho\right]\right]
-i\left[Z_1,\hi\right]=0
& , &
\left[C_2,\ho\right]=0 \label{second}
\\
& \vdots & \nonumber
\end{eqnarray}
where we have taken into account also the additional constraint
$[C(\lambda),\ho]=0$. The generic term, after the first one, in this infinite
sequence of equations is easily shown to be the following:
\begin{eqnarray*}
C_n
\!\! & - & \!\!
\sum_{m=1}^{n}\, \frac{i^m}{m!} \!\!\!
\sum_{\ \ \ k_1+\cdots +k_m=n}
\left[Z_{k_1},\left[\ldots ,\left[Z_{k_m},\ho\right]\ldots\right]\right]
\\
\!\! & - & \!\!
\sum_{m=1}^{n-1}\, \frac{i^m}{m!} \!\!\!
\sum_{\ \ \ k_1+\cdots +k_m=n-1}
\left[Z_{k_1},\left[\ldots,\left[Z_{k_m},\hi\right]\ldots\right]\right]=0,
\end{eqnarray*}
\[
\left[C_n,\ho\right]=0 \ \ \ \ \ \ n=2,3,\ldots\ .
\]
The infinite set of equations can be solved recursively and the solution,
as already anticipated, is not unique. The first equation,
together with the first constraint,
determines $Z_1$ up to an operator commuting with $\ho$ and
$C_1$ uniquely. Indeed, since
\begin{equation}
C_1=\sum_{m=0}^{\infty} P_m\,C_1\,P_m \ \ \ \ \mbox{and} \ \ \ \
[Z_1,\ho]=\sum_{j\neq k} \left(E_k-E_j\right)P_j\,Z_1\,P_k,
\end{equation}
we conclude that
\begin{equation} \label{formulac}
C_1 = \sum_{m=0}^{\infty} P_m\,\hi\, P_m
\end{equation}
and
\begin{equation}
Z_1= \sum_{m=0}^{\infty} P_m\,Z_1\, P_m +
i\sum_{j\neq k} \left(E_k-E_j\right)^{-1} P_j\,\hi\,P_k.
\end{equation}
This last equation admits a `minimal solution' which is obtained by imposing
a further condition, namely
\[
P_m\, Z_1\, P_m =0 \ \ \ \ m=0,1,\ldots\ .
\]
For $n>1$, we will use an analogous reasoning.
Indeed, let us define, for $n\ge 2$, the following operator function:
\begin{eqnarray}
G_n(Z_1,\ldots ,Z_{n-1})
\!\! & := & \!\!
\sum_{m=2}^{n}\, \frac{i^m}{m!} \!\!\!
\sum_{\ \ \ k_1+\cdots +k_m=n} \!\!\!
\left[Z_{k_1},\left[\ldots ,\left[Z_{k_m},\ho\right]\ldots\right]\right]
\nonumber \\
\!\! & + & \!\!
\sum_{m=1}^{n-1}\, \frac{i^m}{m!} \!\!\!
\sum_{\ \ \ k_1+\cdots +k_m=n-1} \!\!\!\!
\left[Z_{k_1},\left[\ldots,\left[Z_{k_m},\hi\right]\ldots\right]\right].
\ \ \ \ \ \label{function}
\end{eqnarray}
Now, assume that the first $n$
equations have been solved. Then, the operator $G_{n+1}(Z_1,\ldots,Z_n)$
is known explicitly and hence
\begin{equation} \label{forc}
C_{n+1}= \sum_{m=0}^{\infty} P_m\ G_{n+1}(Z_1,\ldots,Z_n)\, P_m,
\end{equation}
\begin{equation} \label{forz}
\left[Z_{n+1},\ho\right]= i\sum_{j\neq k} P_j\,
G_{n+1}(Z_1,\ldots,Z_n)\, P_k\ .
\end{equation}
Again, this last equation determines $Z_{n+1}$ up to an operator
commuting with $\ho$.
In general, the choice of a particular solution for $Z_{n+1}$ will also
influence the form of $C_{n+2},Z_{n+2},\ldots\ $.\\
Thus,
we conclude that the sequence of equations defined above
admits infinite solutions
(even in the case when $\ho$ has a nondegenerate spectrum). This fact
had to be expected as a consequence of the non-unicity of $U(\lambda)$.
Anyway,
there is a unique `minimal solution' $\{C_1,Z_1,\ldots\}$
which fulfills the following
additional condition:
\begin{equation}
P_m\, Z_k\, P_m=0\ \ \ \ m=0,1,\ldots\ ,\ k=1,2,\ldots\ .
\end{equation}

The preceding scheme can be extended to the general case when the
interaction component $\hi(\lambda)$ of $\hl$
does not depend linearly on $\lambda$:
\begin{equation}
\hl=\ho+\hi(\lambda)=\sum_{m=0}^\infty \lambda^m\, \hm.
\end{equation}
Indeed, given an operator $X$, let us set
$F_{0}(X)\equiv X$ and
\begin{equation}
F_{n}(X; Z_1,\ldots ,Z_{n}) :=
\sum_{m=1}^{n}\, \frac{i^m}{m!} \!\!\!
\sum_{\ \ \ k_1+\cdots +k_m=n} \!\!\!
\left[Z_{k_1},\left[\ldots
,\left[Z_{k_m}, X\right]\ldots\right]\right],
\end{equation}
for $n\ge 1$. Then we can define the operator function
\begin{eqnarray}
G_n(\ho,\ldots ,\hn ;Z_1,\ldots ,Z_{n-1})
\!\! & := & \!\!
\sum_{m=0}^{n} F_{n-m}(\hm;Z_1,\ldots ,Z_{n-m})
\nonumber \\
\!\! & - &
i[Z_n,\ho] \ \ \ \ \ \ n\ge 1,
\end{eqnarray}
which generalizes definition~{(\ref{function})}. At this point, one can show
that this time the decomposition formula~{(\ref{decomposition})} leads to
the following sequence of equations:
\begin{eqnarray} \label{general}
C_n - i\left[Z_n,\ho \right]
\!\! & = &  \!\!
G_n (\ho,\ldots ,\hn; Z_1,\ldots, Z_{n-1}),
\nonumber \\
\left[C_n, \ho\right]
\!\! & = & \!\!
0 \ \ \ \ \ \ \ \ n\in\mathbb{N}.
\end{eqnarray}
Again,
the general solution of this set of equations can be obtained recursively
by formulae~{(\ref{forc})} and~{(\ref{forz})}.
We want to
show next that it is possible to give simple explicit expressions for
the operators $C_n$, $Z_n$ which do not involve
the eigenprojectors of $\ho$.\\
To this aim, let us denote by $\mathcal{H}_0,\mathcal{H}_1,\ldots$ the
eigenspaces associated with the eigenvalues $E_0,E_1,\ldots$; namely,
let us set:
\[
\mathcal{H}_n\equiv\mathrm{Ran}(P_n),\ \ \
d_n\equiv\mathrm{dim}(\mathcal{H}_n) \ \ \ \
n=0,1,\ldots\ ,\ \ \
d_{-1}\equiv 0.
\]
Now, let $\{|n\rangle\, :\ n=0,1,\ldots\}$ be an orthonormal basis formed
by eigenvectors of $\ho$ such that
\[
\mathcal{H}_n=\mathrm{Span}(|c_n\rangle,\ldots,|c_n+d_n\rangle\}, \ \ \
c_n\equiv\sum_{k=-1}^{n-1} d_n.
\]
We will denote by $A,A^\dagger,\hat{N}$ respectively the annihilation,
creation and number operators relative to this basis:
\[
A^\dagger \left|n\right\rangle =\sqrt{n+1}\, \left|n+1\right\rangle, \ \ \
\hat{N}\left|n\right\rangle=n\left|n\right\rangle.
\]
Then we can write
\begin{equation}
\ho=E_{\hat{N}}
\end{equation}
and express the operators
$G_1\equiv\mathfrak{H}_1$,
$G_n\equiv G_n(\ho,\ldots,\mathfrak{H}_n;Z_1,\ldots,Z_{n-1})$, $n\ge 2$,
in the following form:
\begin{equation}
G_n =g_{\,n}^{[0]}(\hat{N})+\sum_{m=1}^{\infty}\left(
g_{\,n}^{[m]}(\hat{N})\,A^m
+ A^{\dagger\, m} g_{\,n}^{[m]}(\hat{N})\right).
\end{equation}
At this point one can check that the operators
\begin{equation} \label{solc}
C_n=g_{\, n}^{[0]}(\hat{N})+\sum_{m=1}^{d-1}\left(
\chi\left(E_{\hat{N}+m}-E_{\hat{N}}\right) g_{\, n}^{[m]} (\hat{N})\,
A^m + \, h.c.\right),
\end{equation}
\begin{equation} \label{solz}
Z_n= Z_{\, n}^{[0]}+i \sum_{m=1}^{\infty}\left(
\gamma\left(E_{\hat{N}+m}-E_{\hat{N}}\right) g_{\, n}^{[m]} (\hat{N})\,
A^m - \, h.c.\right)
\end{equation}
---
where $d\equiv\sup\{\, d_n : n=0,1,\ldots\}$,
$Z_{\,n}^{[0]}$ is a hermitian operator such that
$[Z_{\,n}^{[0]},\ho]=0$ and the functions $\chi ,\gamma$ are defined by
\begin{eqnarray*}
\chi(0)=1,
&  &
\chi(x)=0 \ \ \ \ x\neq 0 ,
\\
\gamma(0)=0,
&  &
\gamma(x)=\frac{1}{x} \ \ \ x\neq 0
\end{eqnarray*}
---
are solutions of eq.~{(\ref{general})}. This can be readily verified
by direct substitution if one observes that
\[
\chi(x)\, x = 0,\ \ \ \gamma(x)\, x = 1-\chi(x).
\]
We remark that if now $\{\left\|n\right\rangle : n=0,1,\ldots\}$ is
{\it any} orthonormal basis of eigenvectors of $\ho$, so that
\[
\ho\left\|n\right\rangle = E(n)\left\|n\right\rangle,
\]
where in general $E(n)\neq E_n$, and $A$, $A^\dagger$, $\hat{N}$ are
respectively the annihilation, creation, and number operator
associated with this basis,
then formulae~{(\ref{solc})}
and~{(\ref{solz})} still apply with the substitutions
\[
E_{\hat{N}}\ \longmapsto\ E(\hat{N}),\ \ \
d-1 \ \longmapsto\ \infty.
\]
We observe explicitly that with these modifications
formulae~{(\ref{solc})} and~{(\ref{solz})} can be applied formally
ignoring the possible degeneracies in the spectrum of $\ho$.


\section{Perturbative analysis: treatment of the BH} \label{treatment}

In this section, we want to apply the perturbative method outlined in
the preceding one to the balanced Hamiltonian $\breve{H}$. More precisely,
for better illustrating the method, we will use, beside $\breve{H}$,
a Hamiltonian which is unitarily (hence physically) equivalent
to $\breve{H}$.\\
Indeed, let us consider the unitary operator
\begin{eqnarray}
T
\!\! & := & \!\!
\exp\left(i\frac{\pi}{2}\,\hat{n}\,\sigma_x\right)
\nonumber \\
\!\! & = & \!\!
\cos\left(\frac{\pi}{2}\,\hat{n}\right) +i\sin\left(\frac{\pi}{2}\,
\hat{n}\right) \sigma_x .
\end{eqnarray}
One checks easily that the following transformation formulae hold:
\begin{eqnarray}
T\, a\, T^\dagger
\!\! & = & \!\!
- i\, a\,\sigma_x,
\\
T\, \sigma_z\, T^\dagger
\!\! & = & \!\!
e^{i\pi\hat{n}}\,\sigma_z,
\\
T\,\sigma_\pm \, T^\dagger
\!\! & = & \!\!
\cos^2\left(\frac{\pi}{2}\,\hat{n}\right)\sigma_\pm
+\sin^2\left(\frac{\pi}{2}\,\hat{n}\right)\sigma_\mp.
\end{eqnarray}
With the aid of this formulae, we find that
\begin{equation}
\hh := T\, \breve{H}\, T^\dagger = \hho + \hhi,
\end{equation}
where:
\begin{equation}
\hho := \nu\,\hat{n} +
\frac{1}{2}\breve{\delta}\, e^{i\pi\hat{n}}\,
\sigma_z + \gi\,\breve{\eta}\, \nu
\end{equation}
and
\begin{eqnarray}
\hhi
\!\! & := & \!\!
\gi\,\nu\left(a+a^\dagger\right)
\\
\!\! & + & \!\!
\gi\,\nu
\sum_{m=1}^\infty \frac{\breve{\eta}^{m}}{m!}
\left(a^{\dagger\,2} - a^2 +1
-m\right)(a^\dagger -a)^{m-1}\, e^{im\pi\hat{n}}
\left(\sigma_- +(-1)^{m}\sigma_+\right).
\nonumber
\end{eqnarray}
Now, if we assume that $\breve{\eta}\ll 1$, we can drastically reduce
the number of terms that need to be considered in the interaction components
of the Hamiltonians $\breve{H}$ and $\hh$; namely, we can set:
\begin{eqnarray}
\label{interaction1}
\breve{H}_\updownarrow
\!\! & \simeq & \!\!
i\,\gi\,\nu\left( a-a^\dagger\right)\left(\sigma_+ +\sigma_- \right) +
\gi\,\breve{\eta}\,\nu\left(a^{\dagger\,2}-a^2\right)\left(\sigma_+ -
\sigma_- \right) \\
\label{interaction2}
\hhi
\!\! & \simeq & \!\!
\gi\,\nu\left(a+a^\dagger\right)+
\gi\,\breve{\eta}\,\nu\left(a^{\dagger\,2}-a^2\right) e^{i\pi\hat{n}}
\left(\sigma_- -\sigma_+\right).
\end{eqnarray}
Moreover, in order to treat $H_\updownarrow$ and $\hhi$
as perturbations, with perturbative parameter
$\gi$, we will also assume that $\gi\ll 1$.
For simplicity, we will set
\begin{eqnarray}
\breve{H}
\!\! & = & \!\!
e^{-i\left(\breve{Z}_1+\breve{Z}_2+\cdots\right)}
\left(\breve{H}_0+\breve{C}_1+\breve{C}_2+\cdots\right)
e^{i\left(\breve{Z}_1+\breve{Z}_2+\cdots\right)},
\label{decomp} \\
\check{H}
\!\! & = & \!\!
e^{-i\left(\check{Z}_1+\check{Z}_2+\cdots\right)}
\left(\check{H}_0+\check{C}_1+\check{C}_2+\cdots\right)
e^{i\left(\check{Z}_1+\check{Z}_2+\cdots\right)}.
\end{eqnarray}
Thus, differently from section~{\ref{analysis}}, the perturbative parameter
$\gi$ will be included in the operators $\breve{C}_n$, $\check{C}_n$,
$\breve{Z}_n$, $\check{Z}_n$.

Our purpose is to perform the
perturbative analysis up to the second order;
precisely, since we want to compute the first and second order corrections
to the energy spectrum and the first order expression of
the evolution operator, we need to
compute the operators $\breve{C}_1$, $\breve{Z}_1$, and $\breve{C}_2$.\\
At this point, it is convenient to distinguish
three cases: when $\breve{\eta}$
is much smaller than $\gi$, when $\breve{\eta}$ is of the same order of
magnitude of $\gi$, when $\breve{\eta}$ is much larger than $\gi$.

Let us consider the case when $\breve{\eta}\ll \gi$.\\
In this case, the second term which appears respectively in the r.h.s.\ of
eq.~{(\ref{interaction1})} and eq.~{(\ref{interaction2})},
due to the presence of the factor
$\gi\,\breve{\eta}\ll \gi^2$, can be skipped in a second order treatment;
hence we can set
\[
\breve{H}_\updownarrow\simeq i\,\gi\,\nu\left(a-a^\dagger\right)
\left(\sigma_+ +\sigma_-\right), \ \ \
\hhi\simeq \gi\,\nu\left(a+a^\dagger\right).
\]
Let us use the Hamiltonian $\hh$ first.
Observe that, within the given approximation, the subspaces $\mathcal{H}_g$,
$\mathcal{H}_e$ of the total Hilbert space $\mathcal{H}$,
$\mathcal{H}_g\oplus\mathcal{H}_e=\mathcal{H}$,
\[
\mathcal{H}_g :=\mathrm{Span}\{\left|n\right\rangle\otimes\left|g
\right\rangle : n=0,1,\ldots\}, \ \
\mathcal{H}_e := \mathrm{Span}\{\left|n\right\rangle\otimes\left|e
\right\rangle : n=0,1,\ldots\},
\]
are invariant subspaces for $\hh$.
Thus, it will be convenient to split the perturbative problem by restricting
to each invariant subspace. In order to do this, let us denote by
$\ag$, $\eng$, $\hh_g$, $\hho^g$, $\hhi^g$ (resp.\ $a_e,\ldots$)
the restriction of the operators $a$, $\hat{n}$, $\hh$,
$\hho$, $\hhi$ to $\mathcal{H}_g$ (resp.\ to $\mathcal{H}_e$).
Then, we have:
\[
\hg=\hho^g +\hhi^g =
\nu\,\eng-\frac{1}{2}\breve{\delta}\, e^{i\pi\eng}+
\gi\,\nu\left(\ag +\agd\right).
\]
At this point, identify the basis $\{\left|n\right\rangle\otimes\left|
g\right\rangle: n=0,1,\ldots\}$ in $\mathcal{H}_g$
with the basis $\{\left\|n\right\rangle:
n=0,1,\ldots\}$ of section~{\ref{analysis}} and the operators
$\ag$, $\eng$ with $A$, $\hat{N}$. Then, applying formula~{(\ref{solc})},
we find:
\[
\check{C}_1^g= \gi\,\nu\,
\chi(\nu+\breve{\delta}\, e^{i\pi\eng})\,\ag +\, h.c.\ .
\]
Hence, $\check{C}_1^g=0$ unless the {\it resonance condition}
$\nu=\breve{\delta}$
is satisfied; in this case, we have:
\[
\check{C}_1^g=\frac{1}{2}\gi\,\nu\left(\ag +\agd\right)+
\frac{1}{2}\gi\,\nu\left(\ag-\agd\right)e^{i\pi\eng}.
\]
In the subspace $\mathcal{H}_e$, arguing as above, we find that
$\check{C}_1^e$
is zero unless $\nu=\breve{\delta}$, in which case:
\[
\check{C}_1^e =\frac{1}{2}\gi\,\nu\left(a_e^{} + a_e^\dagger \right) -
\frac{1}{2}\gi\,\nu\left(a_e^{}-a_e^\dagger\right)e^{i\pi\hat{n}_e}.
\]
Hence, considering the total Hilbert space $\mathcal{H}$,
at the first perturbative order in $\gi$ we find that
\begin{eqnarray}
\check{C}_1
\!\! & = & \!\! 0 \ \ \ \ \ \mbox{for}\ \nu\neq\breve{\delta},
\\
\check{C}_1
\!\! & = & \!\!
\frac{1}{2}\gi\,\nu\left(a + a^\dagger \right) -
\frac{1}{2}\gi\,\nu\left(a -a^\dagger\right)e^{i\pi\hat{n}}\,\sigma_z
\ \ \ \ \ \mbox{for}\ \nu=\breve{\delta}.
\end{eqnarray}
Moreover, one can check that the minimal solution for
$\check{Z}_1$ is given by:
\begin{eqnarray}
\check{Z}_1
\!\! & = & \!\!
\frac{i}{2}\gi\,\nu\,\gamma(\nu-\breve{\delta})
\left(\left(a - a^\dagger \right) -
\left(a +a^\dagger\right)e^{i\pi\hat{n}}\,\sigma_z\right)
\nonumber \\
\!\! & + & \!\!
\frac{i}{2}\gi\,\frac{\nu}{\nu+\breve{\delta}}
\left(\left(a- a^\dagger \right)
+\left(a+ a^\dagger \right)e^{i\pi\hat{n}}\,\sigma_z\right).
\end{eqnarray}
The operator $\check{Z}_1$ allows us to compute the second order
constant of the motion $\check{C}_2$. The result is found to be:
\begin{eqnarray}
\check{C}_2
\!\! & = & \!\!
\gi^2\,\frac{\nu^2}{\nu+\breve{\delta}}\left(\left(\hat{n}+\frac{1}{2}\right)
e^{i\pi\hat{n}}\,\sigma_z - \frac{1}{2}\right)
\nonumber \\
\!\! & - & \!\!
\gi^2\,\nu^2\,\gamma(\nu-\breve{\delta})
\left(\left(\hat{n}+\frac{1}{2}\right)e^{i\pi\hat{n}}\,\sigma_z +
\frac{1}{2}\right).
\end{eqnarray}
On the other hand, in the `reference frame' associated
with $\breve{H}$, at the first order we have:
\begin{eqnarray}
\label{forca}
\breve{C}_1
\!\! & = & \!\! T^\dagger \check{C}_1\, T =
0 \ \ \ \ \ \mbox{for}\ \nu\neq\breve{\delta},
\\  \label{forcb}
\breve{C}_1
\!\! & = & \!\!
i\,\gi\,\nu\left(a\, \sigma_+ - a^\dagger\,\sigma_- \right)
\ \ \ \ \ \mbox{for}\ \nu=\breve{\delta}.
\end{eqnarray}
Notice that,
in correspondence to the resonance condition $\nu=\breve{\delta}$,
$\breve{C}_1$ coincides with the constant of the motion
$\widehat{\mathcal{S}}$, with
$\omega\equiv\breve{\delta}$, of the Hamiltonian
$\widehat{H}_\mathrm{JC}$ which is unitarily equivalent
to the Jaynes-Cummings Hamiltonian (see section~{\ref{classical}}) and hence
with the prescription of the RWA (see section~{\ref{ion}}).
We could have obtained this result using directly
the  Hamiltonian $\breve{H}$, observing that
$m\,\nu=\breve{\delta},\ m=1,2,\ldots$, is the degeneracy condition for
$\breve{H}_0$ and then applying formula~{(\ref{formulac})}.\\
Thus, it could seem that, in the resonant regime $\nu=\breve{\delta}$,
the first order approximation of $\breve{H}$ coincides with the
RWA Hamiltonian $\breve{H}_0+\breve{C}_1$. Actually, it would be so
if $\breve{Z}_1$ was identically zero (recall formula~{(\ref{decomp})}),
but we have:
\begin{eqnarray}
\breve{Z}_1
\!\! & = & \!\! T^\dagger \check{Z}_1 \, T
\nonumber\\ \label{forza}
\!\! & = & \!\!
-\gi\,\nu\,\gamma(\nu-\breve{\delta})
\left(a\,\sigma_+ + a^\dagger\,\sigma_- \right) -
\gi\,\frac{\nu}{\nu+\breve{\delta}}
\left(a\,\sigma_- + a^\dagger\,\sigma_+ \right) .
\end{eqnarray}
We will see the effects of this fact
in section~{\ref{t-evolution}}.\\
At the second order, we find:
\begin{eqnarray}
\!\!\!\! \breve{C}_2
\!\! & = & \!\!
T^\dagger \check{C}_2\, T
\nonumber \\ \label{formc2}
\!\! & = & \!\!
\frac{\gi^2\,\nu^2}{\nu+\breve{\delta}}\left(\left(\hat{n}+\frac{1}{2}\right)
\sigma_z - \frac{1}{2}\right)
- \gi^2\,\nu^2\,\gamma(\nu-\breve{\delta})
\left(\left(\hat{n}+\frac{1}{2}\right)\sigma_z +
\frac{1}{2}\right).
\end{eqnarray}

Let us now consider the case when $\breve{\eta}\sim\gi$.\\
In this case, since $\gi\,\breve{\eta}\sim\gi^2$, we have to
consider the second term in the r.h.s.\ respectively of
eq.~{(\ref{interaction1})} and eq.~{(\ref{interaction2})}
at the second perturbative order. In this case, one can work directly
with the Hamiltonian $\breve{H}$. Then, the result is the following.
At the first order, the constant of the motion $\breve{C}_1$ is still
given by formulae~{(\ref{forca})} and~{(\ref{forcb})}, $\breve{Z}_1$
by formula~{(\ref{forza})}.
At the second order, one finds:
\begin{eqnarray}
\breve{C}_2
\!\! & = & \!\!
\gi^2\,\frac{\nu^2}{\nu+\breve{\delta}}\left(\left(\hat{n}+\frac{1}{2}\right)
\sigma_z - \frac{1}{2}\right)
-
\gi^2\,\nu^2\,\gamma(\nu-\breve{\delta})
\left(\left(\hat{n}+\frac{1}{2}\right)\sigma_z +
\frac{1}{2}\right)
\nonumber \\ \label{formc2b}
\!\! & - & \!\!
\gi\,\breve{\eta}\,\nu\,\chi(2\nu-\breve{\delta})
\left(a^2\,\sigma_+ + a^{\dagger\,2}\,\sigma_-\right).
\end{eqnarray}
Hence an extra term, with respect to formula~{(\ref{formc2})}, appears
in the expression of $\breve{C}_2$. This term is related to the resonance
$2\nu=\breve{\delta}$.

Eventually, let us consider the case when $\breve{\eta}\gg\gi$.\\
In this case, we have:
\[
\gi^2\ll\gi\breve{\eta}\ll\gi\ll\breve{\eta}\ll 1.
\]
Hence, the term proportional to $\gi\,\breve{\eta}$ can now be regarded as
the leading term at the second perturbative order.
Then, the operators $\breve{C}_1$,
$\breve{Z}_1$ and $\breve{C}_2$ are given again by formulae~{(\ref{forca})},
(\ref{forcb}), (\ref{forza}) and~{(\ref{formc2b})}.

A special attention is deserved by the nearly resonant regime. Indeed, if
the condition
$|\nu -\breve{\delta}|\ll \nu$ is satisfied,
it is convenient to set $\breve{H}= \breve{H}_0^\prime +
\breve{H}_\updownarrow^\prime$,
where:
\begin{eqnarray}
\breve{H}_0^\prime
\!\! & = & \!\!
\nu\left(\hat{n}+\frac{1}{2}\,\sigma_z\right)+\gi\,\breve{\eta}\,\nu,
\\
\breve{H}_\updownarrow^\prime
\!\! & = & \!\!
\frac{1}{2}\left(\breve{\delta}-\nu\right)\sigma_z
\nonumber \\
& + & \!\!
i\,\gi\,\nu\left( a-a^\dagger\right)\left(\sigma_+ +\sigma_- \right) +
\gi\,\breve{\eta}\,\nu\left(a^{\dagger\,2}-a^2\right)\left(\sigma_+ -
\sigma_- \right).
\end{eqnarray}
Thus, the zeroth order Hamiltonian $\breve{H}_0^\prime$ has degeneracies
in its spectrum as in the resonant case and an extra term appears in the
perturbation. This strategy, which is analogous to carefully choosing
the origin of the power expansion of an analytic function,
allows to obtain larger convergence radii
for our pertubative expansions. In the nearly resonant case, one finds that
the expressions of the operators
$\breve{C}_1$, $\breve{Z}_1$ and $\breve{C}_2$ are
the following:
\begin{equation} \label{aaa}
\breve{C}_1 = \frac{1}{2}\left(\breve{\delta}-\nu\right)\sigma_z+
i\,\gi\,\nu\left(a\, \sigma_+ - a^\dagger\,\sigma_- \right),
\end{equation}
\begin{equation}
\breve{Z}_1 =
-\frac{1}{2}\,\gi
\left(a\,\sigma_- + a^\dagger\,\sigma_+ \right),
\end{equation}
\begin{equation} \label{bbb}
\breve{C}_2 = \frac{1}{2}\,
\gi^2 \nu \left(\left(\hat{n}+\frac{1}{2}\right)
\sigma_z - \frac{1}{2}\right).
\end{equation}
Notice that, as in the exactly resonant regime, the operator
$\breve{H}_0^\prime+\breve{C}_1$ coincides with the RWA Hamiltonian (now
obtained using as reference Hamiltonian $\breve{H}_0^\prime$), but
again, since $\breve{Z}_1\neq 0$, it is not the correct first order
approximation of $\breve{H}$.


\section{Corrections to eigenvalues: the RWA and the Bloch-Siegert
shift} \label{spectrum}

By means of
the perturbative method discussed in the previous sections one can
obtain approximate expressions, at each perturbarive order,
of the eigenvalues, the eigenprojectors and of the
evolution operator associated with the BH.\\
For instance, recalling formulae~{(\ref{aaa})}
and~{(\ref{bbb})}, one finds that approximate
expressions, at the second perturbative order,
of the energy levels of the BH
in the nearly resonant regime, i.e.\
for $|\nu-\breve{\delta}|\ll\nu$, are given by the eigenvalues
of the hermitian operator
\begin{eqnarray*}
\mathfrak{h}^{(2)}
\!\! & = & \!\!
\nu\left(\hat{n}+\frac{1}{2}\,\sigma_z\right)
+ \frac{1}{2}\left(\breve{\delta}-\nu\right)\sigma_z
\\
\!\! & + & \!\!
i\,\gi\,\nu\left(a\,\sigma_+ -a^\dagger\,\sigma_-\right)
+\frac{1}{2}\,\gi^2\nu
\left(\left(\hat{n}+\frac{1}{2}\right)\sigma_z-\frac{1}{2}\right).
\end{eqnarray*}
Observe that the eigenspaces of the operator $\mathfrak{h}^{(2)}$ are
the one-dimensional subspace
$
\mathcal{H}_0=\mathrm{Span}\{|0\rangle\otimes|g\rangle\}
$
and the two-dimensional subspaces
\[
\mathcal{H}_{n}=\mathrm{Span}\{|n-1\rangle\otimes|e\rangle, \ \
|n\rangle\otimes|g\rangle\},\ \ \ n=1,2,\ldots\ .
\]
Hence, all one has to do is to diagonalize the
matrix representation of $\mathfrak{h}^{2)}$ in each
of these two-dimensional subspaces; namely, one has to diagonalize the
$2\times 2$ matrices
\begin{equation}
\left[
\begin{array}{cc}
\mathfrak{A}_n + \mathfrak{B}_n
& i\,\gi\,\nu\,\sqrt{n}
\\
-i\,\gi\,\nu\,\sqrt{n} & \mathfrak{A}_n - \mathfrak{B}_n
\end{array}
\right],\ \ \ \ n=1,2,\ldots\ ,
\end{equation}
where:
\begin{eqnarray}
\mathfrak{A}_n
\!\! & = & \!\!
\nu\left(n-\frac{1}{2}\right)
+\gi\,\breve{\eta}\,\nu-\frac{1}{2}\, \gi^2\nu,
\\
\mathfrak{B}_n
\!\! & = & \!\!
\frac{1}{2}\left(\breve{\delta}-\nu\right)+\frac{1}{2}\,\gi^2\nu\, n .
\end{eqnarray}
Thus, the energy levels of the BH $\breve{H}$, in the nearly resonant regime,
are given by
\[
E_0 \simeq -\frac{1}{2}\,\nu + \gi\,\breve{\eta}\,\nu -
\frac{1}{2}\,\gi^2 \nu,
\]
\begin{equation}
E_{n,\pm} \simeq \nu\left(n-\frac{1}{2}\right)+\gi\,\breve{\eta}\,\nu \pm
\sqrt{\frac{1}{4}\left((\breve{\delta}-\nu)+
\gi^2\nu\, n\right)^2 +\gi^2\nu^2 n}-\frac{1}{2}\,\gi^2\nu,
\end{equation}
with $n=1,2,\ldots\ $. The previous expressions coincide, skipping the
second order corrections,
with the ones obtained applying the RWA. This is due to the
fact that, as already observed in section~{\ref{treatment}},
the operator
$\breve{H}_0 + i\,\gi\,\nu\left(a\,\sigma_+ +a^\dagger\sigma_-\right)$
coincides with the result of the application of the RWA to the
Hamiltonian $\breve{H}$, in the
nearly resonant regime $|\nu-\breve{\delta}|\ll\nu$.
Hence, on the one hand,
the RWA gives the correct first order expressions for the eigenvalues.
In the next section we will show that, on the other hand,
the RWA does not give
the correct first order approximation of the evolution operator
associated with the BH (hence, with the ITH).

There is also a second order effect that cannot be predicted if one
simply applies the RWA.
This effect may be compared with the so called `Bloch-Siegert shift'.
In order to clarify this point, let us consider the matrix
representation of a hermitian operator $\mathfrak{h}$
(which can be thought as a perturbative approximation of the Hamiltonian
of a physical system)
in a two-dimensional invariant subspace, with respect to
an orthonormal basis $\{|1\rangle, |2\rangle\}$ in this space:
\[
\mathbf{M}=
\left[
\begin{array}{cc}
\mathfrak{h}_{11} & \mathfrak{h}_{12}
\\
\mathfrak{h}_{21} & \mathfrak{h}_{22}
\end{array}
\right],\ \ \ \ \mathfrak{h}_{11}=\mathfrak{h}_{11}^\ast,\
\mathfrak{h}_{22}=\mathfrak{h}_{22}^\ast,\ \mathfrak{h}_{12}=
\mathfrak{h}_{21}^\ast.
\]
Now, if one sets
\[
\mathfrak{A}=\frac{\mathfrak{h}_{11}+\mathfrak{h}_{22}}{2},\ \ \
\mathfrak{b}=\frac{\mathfrak{h}_{11}-\mathfrak{h}_{22}}{2},\ \ \
\mathfrak{c}=\mathfrak{h}_{12}=\mathfrak{h}_{21}^\ast,
\]
the matrix $\mathbf{M}$ can be rewritten as
\[
\mathbf{M}=
\left[
\begin{array}{cc}
\mathfrak{A}+\mathfrak{b} & \mathfrak{c}
\\
\mathfrak{c}^\ast & \mathfrak{A}-\mathfrak{b}
\end{array}
\right].
\]
and the eigenvalues of the hermitian matrix $\mathbf{M}$ are given by
the following simple formula:
\[
E_\pm(\mathfrak{b})=\mathfrak{A}\pm\sqrt{\mathfrak{b}^2+|\mathfrak{c}|^2}.
\]
Let us assume that $\mathfrak{c}\neq 0$. Then the vectors
$|1\rangle ,|2\rangle$ are not eigenvectors of $\mathfrak{h}$ and
the graphics of the functions $\mathfrak{b}\mapsto E_\pm(\mathfrak{b})$
are the two branches of a hyperbola whose asymptotes intersect at the
point of coordinates $(0,\mathfrak{A})$.
Thus the difference between the two eigenvalues $E_+(\mathfrak{b})-
E_-(\mathfrak{b})$ attains its minimum (`level anticrossing')
at $\mathfrak{b}=0$. This is
also the condition for which the transition probability
$P_{1\!\rightarrow 2}(t)$ assumes periodically the value 1
(otherwise $P_{1\!\rightarrow 2}(t)<1$). Indeed, according to a well known
formula, we have:
\[
P_{1\!\rightarrow 2}(t):= |\langle 2| \exp(-i\,\mathfrak{h}\,t)
|1\rangle|^2=\frac{|\mathfrak{c}|^2}{\mathfrak{b}^2+|\mathfrak{c}|^2}\,
\sin^2\!\left(\sqrt{\mathfrak{b}^2+|\mathfrak{c}|^2}\, t\right).
\]
At this point, suppose that --- due, for instance, to (higher order)
perturbative corrections --- the matrix $\mathbf{M}$ undergoes a
modification of the type
\[
\mathbf{M}\ \ \ \longmapsto \ \ \
\mathbf{M}+ \left[
\begin{array}{lr}
\epsilon & 0 \\
0 & -\epsilon
\end{array}
\right].
\]
Then the level anticrossing condition undergoes a shift:
$\mathfrak{b}+\epsilon=0$. In our case, we can do the following identifications:
\[
\mathfrak{A}\equiv\mathfrak{A}_n,\ \mathfrak{b}\equiv\frac{1}{2}
\left(\breve{\delta}-\nu\right),\ \epsilon\equiv\epsilon_n\equiv
\frac{1}{2}\,\gi^2\nu\, n,\ \ \ n=1,2,\ldots\ .
\]
Thus, the second order level anticrossing shift
for the subspace $\mathcal{H}_n$ is given by
\[
\frac{1}{2}\,\gi^2\nu\, n= \frac{1}{2}\,\frac{\Omega_R}{\sqrt{4\Omega_R^2+
\delta^2}}\,\eta^2\nu\, n,\ \ \ \ n=1,2,\ldots\ .
\]
A similar phenomenon appears in the classical work of Bloch and
Siegert~\cite{Bloch} on the magnetic resonance, whose Hamiltonian
can be replaced, using Floquet's theorem, by a time-independent
effective Hamiltonian containing virtual terms, as shown later by
Shirley~\cite{Shirley}. The presence of these virtual terms gives rise,
at the second perturbative order, to a level anticrossing shift,
which translates into a shift of the magnetic resonance condition,
the Bloch-Siegert shift.


\section{The evolution operator}
\label{t-evolution}

From this point onwards,
for the sake of conciseness, we will use
the following notation. Given a couple of functions $f$ and $h$ of the
perurbative parameter $\gi$, if
$f(\gi)=h(\gi) +\underset{\gi\rightarrow 0}{\mathcal{O}}(\gi^2)$,
we will write simply
\[
f(\gi)\app h(\gi).
\]
Then, let $\mathfrak{U}(t,t_0)$ be the evolution operator associated with
the ion trap Hamiltonian $H(t)$:
\begin{equation}
\left(i\frac{d}{dt}\, \mathfrak{U}\right)(t,t_0)=H(t)\,\mathfrak{U}(t,t_0),\
\ \ \mathfrak{U}(t_0,t_0)=\mathrm{Id}.
\end{equation}
As we have seen, $\mathfrak{U}(t,t_0)$ can be decomposed as
\begin{equation}
\mathfrak{U}(t,t_0)=R_t^\dagger\, T_\Delta^\dagger\,
e^{-i\breve{H}(t-t_0)}\, T_\Delta.
\end{equation}
Moreover, the evolution operator associated with
$\breve{H}$, namely $\evol(t)=e^{-i\breve{H}t}$, admits a perturbative
decomposition. At the first order, we have:
\begin{eqnarray}
\evol(t)
\!\! & \app & \!\!
\exp\left( -i\,
e^{-i\breve{Z}_1} \left( \breve{H}_0+\breve{C}_1 \right)
e^{i\breve{Z}_1}\, t \right)
\nonumber \\
& = & \!\!
e^{-i\breve{Z}_1}\,e^{-i(\breve{H}_0+\breve{C}_1)t}\,e^{i\breve{Z}_1}
\nonumber \\
& = &  \!\!
e^{-i\breve{Z}_1}\, e^{-i\breve{H}_0 t}\,
e^{-i\breve{C}_1 t}\, e^{i\breve{Z}_1}=: \evol_1(t),
\end{eqnarray}
where for obtaining the third line
we have used the fact that $[\breve{C}_1,\breve{H}_0]=0$.
In order to deal with simpler formulae,
let us consider the case when the resonance condition
$\nu=\breve{\delta}$ is exactly satisfied.
In this case, we have that $\breve{H}_0=\nu\,(\hat{n}+\frac{1}{2}\,\sigma_z)$
and $\breve{C}_1=i\,\gi\,\nu\left(a\,\sigma_+ - a^\dagger\,\sigma_-\right)$.
Thus, $\breve{C}_1$ is, up to a unitary transformation,
the infinitesimal generator of the Jaynes-Cummings evolutor $\jc(t)$;
hence:
\begin{eqnarray}
\jch(t)
\!\! & := & \!\!
\exp\left(-i\breve{C}_1\, t\right)
\nonumber \\
& = & \!\!
e^{-i\frac{\pi}{2}\hat{n}}\,\jc(t)\, e^{i\frac{\pi}{2}\hat{n}}
\nonumber \\
& = & \!\!
\left[
\begin{array}{cc}
\cos\left(\gi\,\nu\,\sqrt{\hat{n}+1}\ t\right)
&
\dfrac{\sin\left(\gi\,\nu\,\sqrt{\hat{n}+1}\ t\right)
}{\sqrt{\hat{n}+1}}\, a
\\
&
\\
-\dfrac{\sin\left(\gi\,\nu\,\sqrt{\hat{n}}\ t\right)
}{\sqrt{\hat{n}}}\, a^\dagger
&
\cos\left(\gi\,\nu\,\sqrt{\hat{n}}\ t\right)
\end{array}\right].
\end{eqnarray}
Then it turns out that the expression of the evolution operator
$\mathfrak{R}(t)$ obtained applying the RWA to $\breve{H}$, namely
\begin{equation}
\mathfrak{R}(t) =
\exp(-i\breve{H}_0 t)\,\jch(t),
\end{equation}
would be correct at the first perturbative order
only if the operator $\breve{Z}_1$ was identically zero.
But this is not the case since, for $\nu=\breve{\delta}$,
it turns out that
$\breve{Z}_1=-\frac{1}{2}\,\gi\left(a\,\sigma_- +a^\dagger\,\sigma_+\right)$
and we have:
\begin{equation} \label{unitrasf}
\exp\left(i\breve{Z}_1\right)=
\left[
\begin{array}{cc}
\cos\left(\frac{1}{2}\,\gi\,\sqrt{\hat{n}}\right)
&
-i\,\dfrac{\sin\left(\frac{1}{2}\,\gi\,\sqrt{\hat{n}}\right)
}{\sqrt{\hat{n}}}\, a^\dagger
\\
&
\\
-i\,\dfrac{\sin\left(\frac{1}{2}\gi\,\sqrt{\hat{n}+1}\right)
}{\sqrt{\hat{n}+1}}\, a
&
\cos\left(\frac{1}{2}\,\gi\,\sqrt{\hat{n}+1}\right)
\end{array}\right].
\end{equation}
This result can be also expressed saying that the RWA neglects the first order
correction to the unperturbed eigenprojectors.

In order to get a more explicit comparison of the correct first order
approximate evolution operator $\evol_1(t)$ with
the RWA evolution operator $\mathfrak{R}(t)$, we proceed as follows.
First, we observe that
\begin{eqnarray*}
\evol_1(t)
\!\! & = & \!\!
e^{-i\breve{Z}_1}\, e^{-i\breve{H}_0 t}\, \jch(t)\, e^{i\breve{Z}_1}
\\
& \app & \!\!
e^{-i\breve{Z}_1}\, e^{-i\breve{H}_0 t}\, e^{i\breve{Z}_1}\,
\jch(t).
\end{eqnarray*}
Then, using formula~{(\ref{unitrasf})}, we find that the operator
$e^{-i\breve{Z}_1}\,e^{-i\breve{H}_0 t}\,e^{i\breve{Z}_1}$ has the following
expression:
\[
\left[\!\!
\begin{array}{cc}
\left(\alpha_\gi(\hat{n}) +
\beta_\gi(\hat{n}) e^{i2\nu t}\right) e^{-i\nu\left(\hat{n}+
\frac{1}{2}\right)t}
&
\kappa_\gi(\hat{n})\,
(1-e^{i2\nu t}) \,e^{-i\nu\left(\hat{n}+\frac{1}{2}\right)t} a^\dagger
\\
\kappa_\gi(\hat{n}\!+\!1)\,
(1-e^{-i2\nu t}) \, e^{-i\nu\left(\hat{n}-\frac{1}{2}\right)t} a
&
\left(\alpha_\gi(\hat{n}\! +\! 1) +
\beta_\gi(\hat{n}\! +\! 1) e^{-i2\nu t}\right)
e^{-i\nu\left(\hat{n}-\frac{1}{2}\right)t}
\end{array}
\!\!\right],
\]
where we have set
\[
\alpha_\gi(\hat{n}):= \cos^2\!
\left(\frac{1}{2}\,\gi\,\sqrt{\hat{n}}\right),\ \ \
\beta_\gi(\hat{n}):= \sin^2\!
\left(\frac{1}{2}\,\gi\,\sqrt{\hat{n}}\right),
\]
\[
\kappa_\gi(\hat{n}):=
-i\,\frac{\cos\!\left(\frac{1}{2}\,\gi\,\sqrt{\hat{n}}\right)
\sin\!\left(\frac{1}{2}\,\gi\,\sqrt{\hat{n}}\right)}{\sqrt{\hat{n}}}.
\]
Thus, due to the fact that $\kappa_\gi(\hat{n})\app -\frac{i}{2}\,
\gi\,\mathrm{Id}$, we have:
\[
e^{-i\breve{Z}_1}\,e^{-i\breve{H}_0 t}\, e^{i\breve{Z}_1}
\app\!\!\!\!\!\!\!\!\!\!\diagup \ \ \
e^{-i\breve{H}_0 t},\ \ \ \ \mbox{for}\ t\neq 0.
\]
It follows that, for $t\neq 0$,
\[
\evol(t)\app \evol_1(t)\app e^{-i\breve{Z}_1}\,e^{-i\breve{H}_0 t}\,
e^{i\breve{Z}_1}\,\breve{\jc}(t)
\app\!\!\!\!\!\!\!\!\!\!\diagup \ \ \
e^{-i\breve{H}_0 t}\,\breve{\jc}(t)=
\mathfrak{R}(t).
\]
This proves that the RWA does not provide, already at the first
perturbative order, the correct approximate expression of the evolution
operator associated with the BH (hence with the ITH).\\
Now, in order to obtain a direct comparison of $\evol(t)$ with
$\mathfrak{R}(t)$, observe that
\begin{eqnarray*}
\evol(t)
\!\! & \app & \!\!
e^{-i\breve{Z}_1}\, e^{-i\breve{H}_0 t}\, e^{i\breve{Z}_1}\,
\jch(t)
\\
& = & \!\!
e^{-i\breve{Z}_1}\left(
e^{-i\breve{H}_0 t} \,
e^{i\breve{Z}_1}\, e^{i\breve{H}_0 t}\right) e^{-i\breve{H}_0 t}
\, \jch(t)
\\
& = & \!\!
e^{-i\breve{Z}_1(0)}\, e^{i\breve{Z}_1(-t)}\, e^{-i\breve{H}_0 t}\,
\jch(t)
\\
& \app & \!\!
e^{-i(\breve{Z}_1(0)-\breve{Z}_1(-t))}\, e^{-i\breve{H}_0 t}\,
\jch(t),
\end{eqnarray*}
where $t\mapsto\breve{Z}_1(t)$ is solution of the Heisenberg equation
\[
\left(\frac{d}{dt} \breve{Z}_1\right)(t)=
-i[\breve{Z}_1(t),\breve{H}_0], \ \ \
\breve{Z}_1(0)=\breve{Z}_1.
\]
Then, since
\[
-i[\breve{Z}_1(-t),\breve{H}_0]=i\,\gi\,\nu\left(a\,\sigma_-\,
e^{i\,2\nu t} +a^\dagger\,\sigma_+\,e^{-i\,2\nu t}
\right) =: \breve{Y}_1(t),
\]
we find the following formula:
\begin{equation}
\evol(t)\app
\exp\!\left(-i\int_0^t \breve{Y}_1(\tau)\ d\tau\right)\,
\mathfrak{R}(t).
\end{equation}
This expression provides a direct relation between $\evol(t)$
and $\mathfrak{R}(t)$. Notice that it contains the integral of an
oscillating function. Anyway, since this integral appears as the
argument of an exponential, we are not led to the erroneous conclusion
that its contribution can be neglected, as it is often incorrectly argued
using a Feynman-Dyson expansion of the interaction picture evolution operator
(see section~{\ref{ion}}).


\section{Discussion}
\label{discussion}

In writing the present paper, the authors had in mind two main
aims:
\begin{itemize}

\item to show that the Hamiltonian of a trapped ion interacting with a
laser field can be studied even if the condition $\Omega_R\ll\nu$
is not satisfied, since this condition is incompatible with
applications that physicists consider to be relevant nowadays, for
instance fast ion trap quantum computers;

\item to show that a rigorous perturbative approach can improve
the results obtained by simply applying the RWA, still preserving
the chance of performing explicit and manageable calculations.
\end{itemize}

With regard to the first point, it has been shown that the study of the
ion trap Hamiltonian can be reduced to the study of a time-independent
effective Hamiltonian (the BH) in which the coupling constant
is scarcely sensitive to the Rabi frequency $\Omega_R$,
hence perfectly fit for our aims.
The resonance condition for the BH has a very simple form:
\[
m\,\nu=\breve{\delta}=\sqrt{4\Omega_R^2 +\delta^2},\ \ \ m=1,2,\ldots\ ,
\]
where we recall that $\delta$ denotes the ion-laser detuning.

With regard to the second point, we have shown that a suitable perturbative
approach allows to write a very powerful perturbative expansion of the evolution
operator of the system. Indeed, using the notation of section~\ref{analysis},
we have:
\begin{eqnarray}
e^{-i\,\mathfrak{H}(\lambda)\,t}
\!\! & = & \!\!
\exp\!\left(-i\left(e^{-i(\lambda\,Z_1+\lambda^2 Z_2+\cdots)}\,
(\mathfrak{H}_0+\lambda\, C_1+\cdots)\,
e^{i(\lambda\,Z_1+\lambda^2 Z_2+\cdots)}\right)t\right)
\nonumber \\
\!\! & = & \!\!
e^{-i(\lambda\,Z_1+\lambda^2Z_2+\cdots)}\,
e^{-i(\mathfrak{H}_0+\lambda\,C_1+\cdots)}\,
e^{i(\lambda\,Z_1+\lambda^2Z_2+\cdots)},
\label{evexp}
\end{eqnarray}
where the operators $C_1,Z_1,C_2,Z_2,\ldots$ can obtained by a
recursive algebraic procedure. Notice that any truncation of this
perturbative expansion is a unitary operator, a very valuable
feature for an approximate expression of the evolution operator.
Our approach, whose general validity goes beyond the argument of
this paper, can be applied successfully to the BH pointing out
two main facts. First, the RWA terms and the counter
rotating terms play different roles (but with the same dignity) in
the perturbative expansion~{(\ref{evexp})}. The former appear in
the time-dependent component of the expansion (the one associated
with the operators $C_1,C_2,\ldots$), while the latter appear in
the time-independent component (associated with the operators
$Z_1,Z_2,\ldots$). Second, already at the first perturbative order
the counter rotating terms, completely neglected by the RWA, give rise to a
correction that can be regarded as a perturbative correction to
the unperturbed eigenprojectors (while, as we have seen,
the first order correction to the
eigenvalues coincides with the prescription of the RWA). In
conclusion, we believe that our approach can provide more accurate
expressions of the evolution operator of the ion trap Hamiltonian
for a wide range of intensities of the driving laser field.


\section*{Acknowledgements}

The main results of this paper were presented by one of the authors
(P.\ Aniello) during the 8-th ICSSUR Conference held in Puebla,
Mexico (9-13 June 2003). He wishes to thank the organizers for the kind
hospitality.



\end{document}